\documentclass[aps,prd,reprint,groupedaddress,showpacs,eqsecnum,floatfix]{revtex4-1}

\usepackage{amsmath}
\usepackage{amssymb}
\usepackage{graphicx}

\begin{document}

\title{Anti-de Sitter relativity}

\author{Ion I. Cot\u aescu\\ {\small \it The West 
                 University of Timi\c soara,}\\
   {\small \it V. P\^ arvan Ave. 4, RO-1900 Timi\c soara, Romania}}

\date{\today}

\begin{abstract}
The relative geodesic motion on $(1+3)$-dimensional anti-de Sitter spacetimes is studied in terms of conserved quantities by adapting  the Nachtmann boosting method created initially for the de Sitter spacetimes.  In this approach the Lorentzian isometriy is derived, relating the coordinates of the local chart of a fixed observer  with the coordinates of a mobile chart considered as the rest frame of a massive mobile object moving on a timelike anti-de Sitter geodesic. The transformation of the conserved quantities is also investigated, constructing thus a complete theory of the relative geodesic motion on anti-de Sitter spacetimes. Some applications are discussed among them the problems of twin paradox and Lorentz contraction are solved.

Pacs: 04.02.-q and 04.02.Jb

\end{abstract}

\maketitle

\section{Introduction}

In general relativity the relativistic covariance guarantees the consistency of the theory under any arbitrary coordinate transformations defined as diffeomorphisms.  In this manner one creates the illusion of an universal symmetry but which is sterile in the sense that this is not able to produce conserved quantities. Another sterile general covariance  is the gauge one  which assures the independence of the physical meaning of the theory on the arbitrary gauge transformations of the local orthogonal  nonholonomic  frames. Thus, the only symmetry which may produce conserved quantities remains the covariance under isometries. For this reason, it deserves to study how the conserved quantities are generated via Noether theorem  and, especially, the physical meaning and role of these quantities in describing the geodesic motion.  

The simplest four-dimensional spacetimes of special or general relativity are  vacuum solutions of the Einstein equations whose geometry is determined only by the value of the cosmological constant $\Lambda$. These are the Minkowski flat spacetime (with $\Lambda=0$),  and the hyperbolic spacetimes,  de Sitter (dS) with $\Lambda>0$ and Anti-de Sitter (AdS) having $\Lambda<0$. All these spacetimes have highest possible isometries \cite{W}  representing thus a good framework for studying the role of the conserved quantities seen as the principal observables of the classical or quantum theory. 

The AdS spacetime is the only maximally symmetric spacetime which does not have space translations \cite{W} since its $\Lambda<0$ produces an attraction of the elastic type such that the geodesic motion is oscillatory around the origins of the central charts (i. e. static and spherically symmetric) with ellipsoidal closed trajectories. This particular behavior was studied  by many authors \cite{Cal,Haw,Grif,P1,P2,L,C1} from long time and is still of actual interest \cite{SG,Biz} such that today we are able to understand the principal features of these spacetimes. On the other hand, the higher dimensional AdS manifolds were intensively studied in the framework of the AdS/CFT correspondence of the string theory \cite{Mal,AdSCFT}.  However, there are still unsolved problems as, for example,  the relative geodesic motion in $(1+3)$-dimensional AdS spacetimes and the transformation of the conserved quantities along geodesics observed in different local charts related through isometries.

For the $(1+3)$-dimensional dS spacetime we solved this problem starting with the analyze of the physical meaning of the conserved quantities and invariants \cite{ES,CGRG} that allowed us to find the Lorenzian isometries involved in the relative motion of the dS relativity \cite{CdS}. However, for the AdS spacetime we do not have a similar theory since we succedeed so far to discuss only the physical meaning of the conserved quantities on the timelike geodesics, pointing out the existence of two conserved vectors that may play a similar role as the Runge-Lenz vector of the Keplerian motion \cite{CAdS}. With this starting point, we would like to continue here the construction of the AdS relativity, focusing on the relative motion for determining the isometries between different local chart considered as fixed or mobile inertial natural frames.

As in the dS case \cite{CdS}, we use here a method which was proposed initialy by Nachtmann for constructing  covariant representations of the de Sitter isometry group \cite{Nach}.  The idea is to introduce the coordinates of the local charts  with the help of  point-dependent isometry transformations which are called usually {\em boosts}. In the dS spacetimes we needed to complete the original Nachtmann boosts with suitable gauge transformations in order to give rise simultaneously to local coordinates and desired conserved quantities \cite{CdS}. Fortunatelly, for the AdS spacetimes we may apply a similar method but without resorting to additional gauge transformations.  By using such boosts we define the natural rest frames of the massive point-like particles that will be seen as  mobile frames when the attached particles are moving on timelike geodesics. In this manner we can identify any mobile frame with the help of the conserved quantities on the geodesic of the particle carrying this frame.  Under such circumstances, we may derive the isometry transformation between the coordinates of a mobile frame and those of a fixed one. Thus we are able to study the relative geodesic motion  in terms of conserved quantities, applying on the AdS spacetime the methods of special relativity or of our dS relativity \cite{CdS}. 

The principal new result presented here is the form of the Lorentzian isometries relating  the coordinates of the moving and fixed natural frames on AdS backgrounds. It is remarkable that these have the same form as in the case of the usual Lorentz transformations of special relativity or of the Lorenzian isometries of the dS relativity  \cite{CdS}, depending in the same manner on a conserved momentum. This may appear as paradoxical as long as on the AdS spacetimes there are no space translations and, consequently, no conserved momentum. Nevertheless, on timelike AdS geodesics, one of the conserved vectors studied in Ref. \cite{CAdS} may take over the role of momentum being interpreted as the momentum of the particle carrying the mobile frame when this is passing through the origin of the fixed frame.  However, the mechanisms that lead to this common form of the Lorenzian isometries  are quite different since the motion on the AdS geodesics is oscillatory while on  Minkowski  or dS spacetimes the timelike geodesics are open.  

For marking out the advantages of our approach, we give examples of simple relativistic effects  that can be solved in terms of conserved quantities on AdS spacetime,
focusing on the time dilation of the twin paradox and Lorentz contraction affecting the measurements performed in the origin of the mobile frame. The general method of investigating the relative motion based on our approach is also discussed writing down the general transformations of coordinates and conserved quantities under Lorenzian isometries. 

We start in the second section  with a short review of the properties of the AdS manifolds as hypeboloids in an embedding five-dimensional pseudo-Euclidean spacetime, introducing the principal local charts we need and studying the AdS isometries. The next section is devoted to the conserved quantities on the AdS geodesics focusing on the timelike ones for which we discuss the physical meaning of these observables. Section IV. is devoted to the AdS relativity based on the method of boosting coordinates that allows us to derive the Lorentzian isometries with different parametrizations.  In the next section,  we solve the twin paradox and Lorentz contraction and discuss the general method of relativity applied to the geodesic motion, giving the general rule of transforming coordinates and conserved quantities. The last section presents some concluding remarks.

\section{Anti-de Sitter isometries}

Let us consider the CAdS spacetime $(M,g)$  defined as the universal covering space of the $(1+3)$-dimensional AdS spacetime which  is a vacuum solution of the Einstein equations with $\Lambda<0$ and negative constant curvature. Thus, the AdS manifold is a hyperboloid of radius $R=\frac{1}{\omega} = \sqrt{-\frac{3}{\Lambda}}$ embedded in the $(2+3)$-dimensional pseudo-Euclidean spacetime $(M^5,\eta^5)$ of Cartesian coordinates $z^A$  (labeled by the indices $A,\,B,...= -1,0,1,2,3$) and metric $\eta^5={\rm diag}(1,1,-1,-1,-1)$. These coordinates are global, corresponding to the pseudo-orthonormal basis $\{\nu_A\}$ of the frame into consideration, whose unit vectors satisfy $\nu_A\cdot\nu_B=\eta^5_{AB}$. Any point $z\in M^5$ is  represented by the five-dimensional vector $z=\nu_A z^A=(z^{-1},z^0,z^1,z^2,z^3)^T$ which  transforms linearly under the gauge group $G(\eta^5)=SO(2,3)$  which leave the metric $\eta^5$ invariant.

\subsection{Coordinates on CAdS spacetimes}

The local charts $\{x\}$,  of coordinates $x^{\mu}$ ($\alpha,...\mu,\nu...=0,1,2,3$), can be introduced on $(M,g)$ giving the set of functions $z^A(x)$ which solve the hyperboloid equation,
\begin{equation}\label{hip}
\eta^5_{AB}z^A(x) z^B(x)=\frac{1}{\omega^2}\,.
\end{equation}
The usual chart  $\{t,\vec{x}\}$ with Cartesian spaces coordinates $x^i$ ($i,j,k,...=1,2,3$) is defined by
\begin{eqnarray}
z^{-1}(x)&=&\chi(r)\cos(\omega t)\,,\nonumber\\
z^0(x)&=&\chi(r)\sin(\omega t)\,,\label{Zx}\\
z^i(x)&=&x^i \,, \nonumber
\end{eqnarray}
where we denote $r=|\vec{x}|$ and $\chi(r)=\sqrt{1+\omega {\vec{x}\,}^2}=\sqrt{1+\omega^2 r^2}$. Hereby one obtains the line element
\begin{eqnarray}
ds^{2}&=&\eta^5_{AB}dz^A(x)dz^B(x)\nonumber\\
&=&\chi(r)^2 dt^{2}-\left[\delta_{ij}-\omega^2 \frac{x^{i}x^{j}}{\chi(r)^2}\right]dx^{i}dx^{j}\,.\label{line1}
\end{eqnarray}
The associated central chart $\{t,r,\theta,\phi\}$ with spherical coordinates, canonically related to the Cartesian ones, $\vec{x}\to (r,\theta,\phi)$, has the  line element 
\begin{equation}\label{line2}
ds^2=\chi(r)^2 dt^2-\frac{dr^2}{\chi(r)^2}-r^2(d\theta^2+\sin ^2 \theta\, d\phi^2)\,.
\end{equation}
The AdS spacetime is covered once by the time $t\in[0,2\pi)$ because of the periodicity of the coordinates $z^{-1}$ and $z^0$ but if we consider that $t\in {\Bbb R}^+$ then we are in its universal covering space, CAdS. Here any motion is returning  periodically in the same state at times $t+n\frac{2\pi}{\omega}$ for all $n\in {\Bbb N}$.  In what follows this fact is a matter of course restricting ourselves to study the motion during only a period. 

Apart from the above usual charts, it is useful to consider the central chart $\{\tilde x\}=\{t,\rho,\theta,\phi\}$ resulted after the substitution \cite{P1,P2}
\begin{equation}\label{rrho}
r=\frac{\rho}{\tilde\chi(\rho)}\,, \quad \tilde\chi(\rho)=\sqrt{1-\omega^2\rho^2}\,,
\end{equation}
where $0\le \rho<\frac{1}{\omega}$. Then the embedding equations become
\begin{eqnarray}
z^{-1}(\tilde x)&=&\frac{1}{\omega\tilde\chi(\rho)}\cos(\omega t)\,,\nonumber\\
z^0(\tilde x)&=&\frac{1}{\omega\tilde\chi(\rho)}\sin(\omega t)\,,\nonumber\\
z^1(\tilde x)&=&\frac{\rho}{\tilde\chi(\rho)}\sin \theta \cos \phi \,,\label{Ztx}\\
z^2(\tilde x)&=&\frac{\rho}{\tilde\chi(\rho)}\sin \theta \sin \phi \,,\nonumber\\
z^3(\tilde x)&=&\frac{\rho}{\tilde\chi(\rho)}\cos \theta\,,\nonumber
\end{eqnarray}
while the line element reads
\begin{equation}\label{line3}
ds^2=\frac{1}{\tilde\chi(\rho)^2}\left[ dt^2-\frac{d\rho^2}{\tilde\chi(\rho)^2}-\rho^2(d\theta^2+\sin ^2 \theta\, d\phi^2)\right]\,.
\end{equation}  
In this chart  the components of the four-velocity are denoted as $\tilde u^{\mu}=\frac{d\tilde x^{\mu}}{ds}$.

\subsection{Isometries}

The CAdS spacetimes are homegeneous spaces of the gauge group $G(\eta^5)=SO(2,3)$ whose transformations leave invariant the metric $\eta^5$ of the embedding manifold $(M^5,\eta^5)$ and implicitly Eq. (\ref{hip}).  For this group we adopt the canonical parametrization
\begin{equation}
{\frak g}(\xi)=\exp\left(-\frac{i}{2}\,\xi^{AB}{\frak S}_{AB}\right)\in SO(2,3) 
\end{equation}
with skew-symmetric parameters, $\xi^{AB}=-\xi^{BA}$,  and the covariant generators of the fundamental representation of the $so(2,3)$ algebra carried by $M^5$ having the matrix elements
\begin{equation}
({\frak S}_{AB})^{C\,\cdot}_{\cdot\,D}=i\left(\delta^C_A\, \eta_{BD}^5
-\delta^C_B\, \eta_{AD}^5\right)\,.
\end{equation}

In any local chart $\{x\}$,  defined by the functions $z=z(x)$, each transformation ${\frak g}\in SO(2,3)$ gives rise to the isometry $x\to x'=\phi_{\frak g}(x)$ derived from the system of equations $z[\phi_{\frak g}(x)]={\frak g}z(x)$. The simplest isometries are the rotations  ${\frak r}\in SO(3)\subset SO(2,3)$ generated by ${\frak J}_i=\frac{1}{2}\varepsilon_{ijk}{\frak S}_{jk}$  that  transform linearly the Cartesian coordinates,   $x^i\to \phi^i_{\frak r}(x)=R_{ij}x^j$, when these are proportional with $z^i$ as in Eq. (\ref{Zx}). Consequently, in this case the $SO(3)$ symmetry becomes {\em global} \cite{ES}. Other rotations in the plane $\{z^{-1}, z^0\}$ form the subgroup $T(1)_H$  generated by ${\frak H}={\frak S}_{-1,0}$ as
\begin{equation}
\exp(-i {\frak H}\alpha)\, : \quad
\begin{array}{lll}
z^{-1}&\to&z^{-1}\cos \alpha-z^0\sin\alpha\\
z^0&\to&z^{-1}\sin \alpha+z^0\cos\alpha\\
z^i&\to&z^i
\end{array}
\end{equation}
producing the time translations $t\to t+\frac{\alpha}{\omega}$. In addition, there are three generators of Lorentz transformations, ${\frak K}_i= {\frak S}_{i0}$, acting on $M^5$. For example, those generated by ${\frak K}_1$,  
\begin{equation}
\exp(-i {\frak K}_1\alpha)\, : \quad
\begin{array}{lll}
z^{-1}&\to&z^{-1}\\
z^0&\to&z^{0}\cosh \alpha+z^1\sinh\alpha\\
z^1&\to&z^{1}\cosh \alpha+z^0\sinh\alpha\\
z^2&\to&z^2\\
z^3&\to&z^3
\end{array}
\end{equation}
give the isometries
\begin{eqnarray}
t&\to& \frac{1}{\omega}\arctan \left(\tan \omega t\cosh \alpha +x^1{\rm sec}\, \omega t\sinh \alpha\right)\,,\nonumber\\
x^1&\to&x^1\cosh\alpha+\chi(r)\sin\omega t\sinh\alpha\,,\nonumber\\
x^2&\to&x^2\,,\nonumber\\
x^3&\to&x^3\,.\nonumber
\end{eqnarray}
The last three generators, ${\frak N}_i={\frak S}_{i,-1}$, are of the Lorentz type but involving the coordinate $z^{-1}$ instead of $z^0$. The transformations along the $z^1$ azis,
\begin{equation}
\exp(-i {\frak N}_1\alpha)\, : \quad
\begin{array}{lll}
z^{-1}&\to&z^{-1}\cosh \alpha+z^1\sinh\alpha\\
z^0&\to&z^{0}\\
z^1&\to&z^{1}\cosh \alpha+z^{-1}\sinh\alpha\\
z^2&\to&z^2\\
z^3&\to&z^3
\end{array}
\end{equation}
give rise to the isometries
\begin{eqnarray}
t&\to& \frac{1}{\omega}{\rm arccot}\, \left({\rm cot}\, \omega t\cosh \alpha +x^1{\rm csc}\, \omega t\sinh \alpha\right)\,,\nonumber\\
x^1&\to&x^1\cosh\alpha+\chi(r)\cos\omega t\sinh\alpha\,,\nonumber\\
x^2&\to&x^2\,,\nonumber\\
x^3&\to&x^3\,.\nonumber
\end{eqnarray}
We specify that the generators $\{ {\frak J}_i, {\frak K}_i\}$ are of the $so(1,3)$ subalgebra while ${\frak H}$ and ${\frak N}_i$ are the specific ones of the $so(2,3)$ algebra.

We presented the above isometries in Cartesian coordinates but these can be rewritten at any time in spherical coordinates by using the above indicated substitutions.

\section{Conserved quantities on timelike geodesics}

In general, after integrating the geodesic equations, one obtains the geodesic trajectories depending on some integration constants that must get a physical interpretation. This is possible only by expressing them in terms of conserved quantities on geodesics.  

\subsection{Conserved quantities}

The conserved quantities are given by the Killing vectors associated to the $SO(2,3)$  isometries  which are defined (up to a multiplicative constant)  as \cite{ES},
\begin{equation}
k_{(AB)\,\mu}=z_A\partial_{\mu} z_B -z_B\partial_{\mu} z_A,,\quad  z_A=\eta^5_{AC}z^C\,,
\end{equation}
In Ref. \cite{CAdS} we introduced ten independent conserved quantities associated to the $SO(2,3)$ generators, for any point-like particle  of mass $m$, freely falling on a CAdS background. By using the four-velocity $u^{\mu}=\frac{dx^{\mu}}{ds}$ we defined, in any chart, the energy
\begin{equation}
{\frak H} \to E=m\omega k_{(-1,0)\,\mu}u^{\mu}\,,
\end{equation}
and the angular momentum  components
\begin{equation}
{\frak J}_i \to L_i=m\frac{1}{2}\varepsilon_{ijk}k_{(j,k)\,\mu}u^{\mu}\,,
\end{equation}
that have the traditional  physical meaning. In addition, there are two more conserved vectors having the components
\begin{eqnarray}
{\frak K}_i \to K_i&=&mk_{(i,0)\,\mu}u^{\mu} \,,\\
{\frak N}_i\to  N_i&=&mk_{(i,-1)\,\mu}u^{\mu} \,.
\end{eqnarray} 
These  conserved quantities represent the components of a skew-symmetric tensor ${\cal K}_{(AB)}=mk_{(AB)\,\mu} u^{\mu}$ that can be written in matrix form as   
\begin{equation}\label{KK}
{\cal K}=
\left(
\begin{array}{ccccc}
0&\frac{E}{\omega}&-N_1&-N_2&-N_3\\
-\frac{E}{\omega}&0&-K_1&-K_2&-K_3\\
N_1&K_1&0& L_3&- L_2\\
N_2&K_2&- L_3&0& L_1\\
N_3&K_3& L_2&- L_1&0
\end{array}\right)\,.
\end{equation}
The transformations ${\frak g}\in SO(2,3)$, generating the  isometries $x\to x'=\phi_{\frak g}(x)$,  transform the conserved quantities according to the rule 
\begin{equation}\label{KAB}
{\cal K}_{(AB)}'={\frak g}_{A\,\cdot}^{\cdot\,C}\,{\frak g}_{B\,\cdot}^{\cdot\,D}\,{\cal K}_{(CD)}\,,
\end{equation}
where  ${\frak g}_{A\,\cdot}^{\cdot\,B}=\eta^5_{AC}\,{\frak g}^{C\,\cdot}_{\cdot \,D}\, \eta^{5\,BD}$ are the matrix elements of the adjoint matrix $\overline{\frak g}=\eta^5\,{\frak g}\,\eta^5$. Consequently,  this transformation can be written simpler as  ${\cal K}'=\overline{\frak g}\,{\cal K}\,\overline{\frak g}^T$. Thus any transformation ${\frak g}\in SO(2,3)$ generates an isometry between two local charts and transforms, simultaneously, all the conserved quantities that can be measured in these charts. 

Now  we can verify that all the conserved quantities carrying space indices ($i,j,...$) transform alike under rotations as $SO(3)$ vectors or tensors. Moreover, the condition  $z^i\propto x^i$ fixes the same (common)  three-dimensional basis $\{\vec{\nu}_1,\vec{\nu}_2,\vec{\nu}_3\}$  in  both the Cartesian charts, of  $M^5$ and respectively $M$.  Then we say that the $SO(3)$ symmetry is {global} \cite{ES}  and we use the vector notation for the conserved quantities as well as for the local Cartesian coordinates on $M$.
However, this basis must not be confused with that of the local orthogonal frames on $M$ which will be defined canonically later.

For studying these conserved quantities on the timelike geodesics  it is convenient to work in the chart $\{t,\rho,\theta,\phi\}$ taking the angular momentum along the third axis,  $\vec{L}=L\vec{\nu}_3=(0,0,L)$, for restricting the motion in the equatorial plane, with $\theta=\frac{\pi}{2}$  and $\tilde u^{\theta}=0$. Then the non-vanishing conserved quantities can be written as 
\begin{eqnarray}
E&=&\frac{m }{\tilde{\chi}^2}\tilde u^t\,,\label{conE}\\
L&=&\frac{m \rho^2 }{\tilde{\chi}^2}\tilde u^{\phi}\,,\label{conL}\\
K_1&=&\frac{m }{\omega\tilde{\chi}^2}\left(-\omega \rho \tilde u^t\cos\omega t\cos\phi \right. \nonumber\\
&&\left.+\tilde u^{\rho}\sin\omega t\cos\phi -\rho \tilde u^{\phi}\sin\omega t\sin\phi\right)\,,
\label{K1}\\
K_2&=&\frac{m }{\omega\tilde{\chi}^2}\left(-\omega \rho \tilde u^t\cos\omega t\sin\phi \right. \nonumber\\
&&\left.+\tilde u^{\rho}\sin\omega t\sin\phi +\rho \tilde u^{\phi}\sin\omega t\cos\phi\right)\,,\\
N_1&=&\frac{m }{\omega\tilde{\chi}^2}\left(\omega \rho \tilde u^t\sin\omega t\cos\phi \right. \nonumber\\
&&\left.+\tilde u^{\rho}\cos\omega t\cos\phi -\rho \tilde u^{\phi}\cos\omega t\sin\phi\right)\,,\\
N_2&=&\frac{m }{\omega\tilde{\chi}^2}\left(\omega \rho \tilde u^t\sin\omega t\sin\phi \right. \nonumber\\
&&\left.+\tilde u^{\rho}\cos\omega t\sin\phi +\rho \tilde u^{\phi}\cos\omega t\cos\phi\right)\,,\label{N2}
\end{eqnarray}
while $K_3=N_3=0$. Hereby we deduce the following obvious properties
\begin{equation}\label{property}
\vec{K}\cdot \vec{L}=\vec{N}\cdot \vec{L}=0\,, \quad \vec{K}\land \vec{N}=-\frac{E}{\omega} \vec{L}\,,
\end{equation}
and verify the identity  
\begin{equation}\label{invariant}
E^2+\omega^2\left({\vec{L}\,}^2 -{\vec{K}\,}^2-{\vec{N}\,}^2\right)=m^2\tilde u^2=m^2\,,
\end{equation}
defining the principal invariant corresponding to the first Casimir operator of the $so(2,3)$ algebra. Notice that in the classical theory the second invariant of this algebra vanishes since there is no spin \cite{CGRG}.

Finally, we must specify that a mobile of mass $m$  can stay at rest in origin, $\vec{x}=0$,  on a world line along the vector field $\partial_t$, only when $E=m$ and $\vec{L}=\vec{K}=\vec{N}=0$. Then, the matrix  (\ref{KK}) takes the simplest form
\begin{equation}
{\cal K}_o=\left(
\begin{array}{ccccc}
0&\frac{m}{\omega}&0&0&0\\
-\frac{m}{\omega}&0&0&0&0\\
0&0&0&0&0\\
0&0&0&0&0\\
0&0&0&0&0
\end{array}\right)\,,
\end{equation}
depending only on the mobile mass. We observe that the group  $SO(3)\otimes T(1)_H \subset SO(1,4)$ is the {stable group} of this matrix since $\overline{\frak g}\,{\cal K}_{o}\,\overline{\frak g}^T= {\cal K}_{o}$ for any transformation ${\frak g}$ of this group. 

\subsection{Timelike geodesics}

In the case of the timelike geodesics we may exploit the identity $\tilde u^2=1$ and Eqs. (\ref{conE}) and (\ref{conL}) for obtaining the radial component 
\begin{equation}\label{ur}
\tilde u^{\rho}=\tilde\chi(\rho)^2\left[\frac{E^2}{m^2} \tilde\chi(\rho)^2+\frac{\omega^2 L^2}{m^2}-\frac{L^2}{m^2\rho^2}-1\right]^{\frac{1}{2}}\,,
\end{equation}
that allows us to derive the following prime integrals \cite{CAdS}),
\begin{eqnarray}
\left(\frac{d\rho}{dt}\right)^2+\omega^2\rho^2+\frac{L^2}{E^2\rho^2}&=&1+\frac{\omega^2 L^2}{E^2}-\frac{m^2}{E^2}\,,\\
\frac{d\phi}{dt}&=&\frac{L}{E\rho^2}\,,
\end{eqnarray}
that give the geodesic equations 
\begin{eqnarray}
\rho(t)&=&\left[\kappa_1+\kappa_2\cos 2\omega (t-t_0)\right]^{\frac{1}{2}}\,,\label{GEO1}\\
\phi(t)&=&\phi_0+{\rm arctan} \left[\sqrt{\frac{\kappa_1-\kappa_2}{\kappa_1+\kappa_2}}\tan\omega(t- t_0)\right],\label{GEO2}
\end{eqnarray}
where
\begin{eqnarray}
\kappa_1&=&\frac{\omega^2 L^2+E^2-m^2}{2\omega^2 E^2}\,,\label{Kap1}\\
\kappa_2&=&\frac{1}{2\omega^2 E^2}\left[(E+m)^2-\omega^2 L^2\right]^{\frac{1}{2}} \left[(E-m)^2-\omega^2 L^2\right]^{\frac{1}{2}}\,,\nonumber\\
\label{Kap2}
\end{eqnarray}
satisfy the identity 
\begin{equation}\label{Kap3}
\kappa_1^2-\kappa_2^2=\frac{L^2}{\omega^2 E^2}\,.
\end{equation}
Thus we solve the geodesic equation in terms of conserved quantities which give a physical meaning to the principal integration constants. The remaining  ones,  $t_0$ and $\phi_0$, determine only the initial position of the mobile and implicitly of its trajectory. 

{ \begin{figure}
    \centering
    \includegraphics[scale=.60]{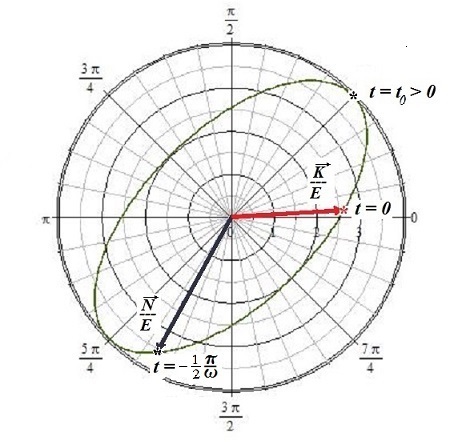}
    \caption{Anti-de Sitter timelike geodesic with arbitrary initial conditions, $\phi_0=\frac{\pi}{4}$ and $t_0=\frac{1}{3}\frac{\pi}{\omega}$. When $t=t_0$ the mobile reaches the aphelion. The vector $\frac{\vec{K}}{E}$ indicates the position of the mobile at $t=0$ while $\frac{\vec{N}}{E}$ gives its position at $t=-\frac{1}{2}\frac{\pi}{\omega}$. }
  \end{figure}}

In this manner, we recover the well-known behavior of the time-like geodesic motion which is oscillatory with frequency $\omega$ having a closed trajectory which has, in general, an ellipsoidal form in the domain $\rho\in [\rho_{\rm min}, \rho_{\rm max}]$ where
\begin{equation}\label{rhomm}
\rho_{\rm min}=\sqrt{\kappa_1-\kappa_2}\,, \quad \rho_{\rm max}=\sqrt{\kappa_1+\kappa_2}\,,
\end{equation}
satisfy $\rho_{\rm min}\le\rho_{\rm max}<\frac{1}{\omega}$ for any values of $E$ and $L$. The trajectory is symmetric with respect the origin $\vec{x}=0$ such that the mobile, moving in trigonometric sense,  reaches two opposite aphelions and perihelions  during a period. From Eq. (\ref{GEO1}) we see that $t_0$  is the time when the mobile is passing through an aphelion while $\phi_0$  is the angle between the axis $\vec{\nu}_1$ and the  major semiaxis of the ellipsoidal trajectory (as in Fig.1). 

 { \begin{figure}
    \centering
    \includegraphics[scale=.60]{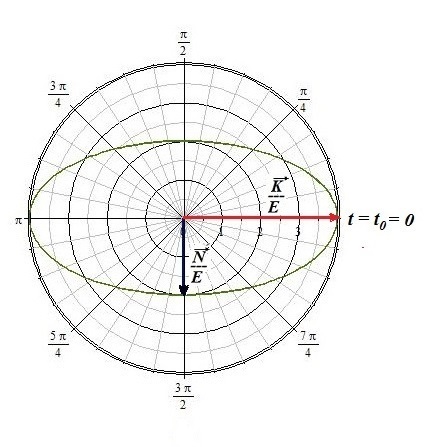}
    \caption{The vectors $\frac{\vec{K}}{E}$  and $\frac{\vec{N}}{E}$  playing the role of Runge-Lenz vectors for  $t_0=0$ ($\phi_0=0$). }
  \end{figure}}

{ \begin{figure}
    \centering
    \includegraphics[scale=.60]{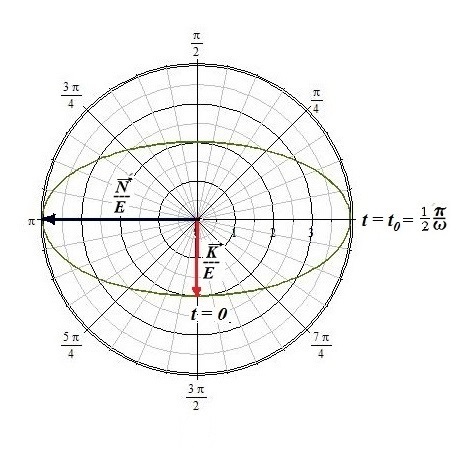}
    \caption{The vectors $\frac{\vec{K}}{E}$  and $\frac{\vec{N}}{E}$ playing the role of Runge-Lenz vectors for   $t_0=\frac{1}{2}\frac{\pi}{\omega}$ ($\phi_0=0$). }
  \end{figure}}

We specify that the axes $\vec{\nu}_1$ and $\vec{\nu}_2$ can be rotated at any time such that the major axis be along the direction of $\vec{\nu}_1$ which means that we may set $\phi_0=0$ without losing generality. Then the geodesic equations can be written easily even in Cartesian space coordinates,  
\begin{eqnarray}
x^1(t)&=&\frac{\rho_{\rm max} }{\sqrt{1-\omega^2 \rho(t)^2}}\,\cos \omega (t-t_0)\,,\label{geo1} \\
x^2(t)&=&\frac{\rho_{\rm min} }{\sqrt{1-\omega^2 \rho(t)^2}}\,\sin \omega (t-t_0)\,,\label{geo2}\\
x^3(t)&=&0\,,\label{geo3}
\end{eqnarray}
as it results from Eqs. (\ref{rrho}), (\ref{GEO1}) and (\ref{GEO2}) for $\phi_0=0$.

Now it remains to analyze the role of the conserved vectors $\vec{K}$ and $\vec{N}$ that depends on $E$ and $L$ as well as on $t_0$ and $\phi_0$. In the simpler case of $\phi_0=0$ their non-vanishing components read
\begin{eqnarray}
K_1&=&E\rho_{\rm max}\cos\omega t_0\,, \label{KER1}\\
K_2&=&-E\rho_{\rm min}\sin\omega t_0\,,\label{KER2}\\
N_1&=&-E\rho_{\rm max}\sin\omega t_0 \,,\label{NER1}\\
N_2&=&-E\rho_{\rm min}\cos\omega t_0\,,\label{NER2}
\end{eqnarray}
complying with the specific properties
\begin{eqnarray}
|\vec{K}|&=&E \left[\kappa_1+\kappa_2\cos 2\omega t_0\right]^{\frac{1}{2}}\,,\label{prop1}\\
|\vec{N}|&=&E \left[\kappa_1-\kappa_2\cos 2\omega t_0\right]^{\frac{1}{2}}\,,\\
\vec{K}\cdot \vec{N}&=&-E^2\kappa_2 \sin 2\omega t_0\,.\label{prop3}
\end{eqnarray}
In the general case of $\phi_0\not=0$ these components have the form given in the Appendix A where we discuss the inverse problem. 

Hereby we understand that the vectors $\frac{\vec{K}}{E}$ and $\frac{\vec{N}}{E}$ are radial,  indicating the mobile positions at $t=0$ and respectively $t=-\frac{1}{2}\frac{\pi}{\omega}$, as in Fig. 1. For $t_0=0$ and $t_0=\frac{1}{2}\frac{\pi}{\omega}$ these vectors become orthogonal, playing the role of Runge-Lenz vectors \cite{CAdS} as in Figs. 2 and 3. More specific, for $t_0=0$ the vector $\frac{\vec{K}}{E}$ lays over the major semiaxis and $\frac{\vec{N}}{E}$ gives the minor one but for $t_0= \frac{1}{2}\frac{\pi}{\omega}$ these vectors change their roles between themselves.

There are two important particular cases.  
(I) When $L=\frac{1}{\omega}(E-m)$ then $\kappa_2=0$ and the motion becomes uniform, $\phi=\phi_0+\omega (t-t_0)$, on a circle of radius $\rho_c=\sqrt{2\kappa_1}=\frac{1}{\omega}\sqrt{\frac{E-m}{E}}$. Then the radial vectors    $\frac{\vec{K}}{E}$ and $\frac{\vec{N}}{E}$, having the same norm $\rho_c$, are orthogonal regardless the value of $t_0$ which determines only their position. (II)  For  $L=0$ we have $\kappa_2=\kappa_1$ such that the oscillation becomes rectilinear, since $\phi=\phi_0$, and harmonic,  with the amplitude $a$,
\begin{equation}\label{ampl}
\rho(t)=a\sin \omega(t-t_0)\,, \quad a=\sqrt{2\kappa_1}=\frac{\sqrt{E^2-m^2}}{\omega E}\,,
\end{equation}
on the direction  $\vec{n}=\vec{\nu}_1\cos\phi_0+\vec{\nu}_2\sin\phi_0$. In this case the vectors $\vec{K}$ and $\vec{N}$ are relevant only for the following particular choices of the initial conditions
\begin{eqnarray}
&&{\rm (IIa):}~~ t_0=0~~~ \to \left\{\begin{array}{l}
\vec{K}=E\rho_{\rm max}\vec{n}\,,\\
\vec{N}=0\,,
\end{array}\right.\\
&&{\rm (IIb):}~~ t_0=\frac{1}{2}\frac{\pi}{\omega} \to \left\{\begin{array}{l}
\vec{K}=0\,,\\
\vec{N}=E\rho_{\rm max} \vec{n}\,.
\end{array}\right. \label{IIb}
\end{eqnarray}
Thus the conserved vectors $\vec{K}$ and $\vec{N}$ are related with the trajectory parameters and the initial condition such that they could be useful in analyzing the CAdS kinematics.

\section{Relativity}

Recently we have studied the relativity on the dS manifolds \cite{CdS} applying the Nachtmann method \cite{Nach} of boosting coordinates which takes over the Wigner theory \cite{Wig} but in configurations instead of momentum representation. This method can be used for the CAdS spacetimes too since these manifolds may be seen as spaces of left cosets $SO(2,3)/L^{\uparrow}_{+}$ where $L^{\uparrow}_{+}$ is the Lorentz group that plays the role of the gauge group of $(M,g)$, leaving invariant the metric $\eta={\rm diag}(1,-1,-1,-1)$ of the Minkowskian pseudo-Euclidean model of this manifold. In what follows we develop this formalism  denoting for brevity $G=G(\eta)=L^{\uparrow}_{+}$ and $G_5=G(\eta^5)=SO(2,3)$. 

\subsection{Boosting coordinates}

Any AdS spacetime can be constructed as a space of left cosets $G_5/G$ starting with the fixed point $z_o=(\omega^{-1},0,0,0,0)^T\in M^5$ (of local coordinates $(0,0,0,0)$).  Then, the whole AdS manifold can be built as the orbit $M=\{{\frak g} z_o |{\frak g}\in G_5/G\} \subset M^5$ since the subgroup $G$ is just the stable group of $z_o$ obeying ${\frak g}z_o=z_o\,,\, \forall {\frak g}\in G$. Consequently, any point $z(x)\in M$ can be reached  performing the boost transformation ${\frak b}(x):\,z_o\to z(x)={\frak b}(x)z_o$ which defines the functions $z^A(x)$ of the local coordinates $\{x\}$. In fact, these boosts are sections in the principal fiber bundle on  $(M,g)\sim  G_5/G$ whose fiber is just the isometry group $G_5$.  Once the AdS spacetime is defined we can take an open time axis for obtaining the CAdS one. 

This formalism offers one, in addition, the advantage of defining  the {\em canonical} five-dimensional 1-forms
\begin{equation}\label{Om}
\hat\omega(x)={\frak b}^{-1}(x)d\,{\frak b}(x)\,z_o
\end{equation}  
whose components 
\begin{equation}
\hat\omega^{\hat\alpha}(x)=\hat e^{\hat\alpha}_{\mu}(x)dx^{\mu}\,,\quad 
\hat\omega^{-1}(x)=0\,,
\end{equation}  
give the canonical gauge fields (or tetrads)  $\hat e^{\hat\mu}$ of the local co-frames associated to the fields $e_{\hat\mu}$  of the orthogonal local frames \cite{Nach}. These fields are  labelled by the local indices $\hat\alpha,\hat\mu,...$ having the same range as the natural ones. They have the duality and orthogonality properties,
\begin{equation}  
e_{\hat\alpha}^{\mu}\hat e_{\nu}^{\hat\alpha}=\delta_{\mu}^{\nu}\,,\quad
e_{\hat\alpha}^{\mu}\hat e_{\mu}^{\hat\beta}=\delta_{\hat\alpha}^{\hat\beta}\,,
\quad 
g_{\mu\nu} e_{\hat\alpha}^{\mu} e^{\nu}_{\hat\beta}=
\eta_{\hat\alpha \hat\beta}\,, 
\end{equation}
giving  the line element of any chart $\{x\}$  as  
\begin{eqnarray}
ds^2&=&\hat\eta_{AB}\hat\omega^A\,\hat\omega^B\nonumber\\
&=&\eta_{\hat\alpha\hat\beta}\hat e^{\hat\alpha}_{\mu}\hat e^{\hat\beta}_{\nu}dx^{\mu}dx^{\nu}=g_{\mu\nu}dx^{\mu}dx^{\nu}.
\end{eqnarray} 
In general, the boosts are defined up to an {\em arbitrary} gauge, ${\frak b}(x)\to {\frak b}(x)\lambda^{-1}(x)$, $\lambda(x)\in G$, that does not affect the functions $z^A(x)$ but changes the gauge fields,  transforming the  1-forms  as $\hat\omega(x) \to \lambda(x)\,\hat\omega(x)$ \cite{Nach}.    

The boost may be chosen for determining  the type of coordinates we desire. For example, the coordinates of the chart $\{t,\vec{x}\}$ with Cartesian space coordinates can be introduced by using the boost
\begin{eqnarray}\label{Cboost}
&&{\frak b}(t,\vec{x})=  \exp\left(-i {\frak H}\,\omega t\right) \nonumber\\
&&~~~~~~~~~\times\exp\left(-i {\frak N}_i \frac{x^i}{r}\,{\rm arctanh}\left(\frac{\omega r}{\chi}\right)\right)=\nonumber\\
&&\left(\begin{array}{ccccc}
\chi\cos\omega t&-\sin\omega t&\omega x^1\cos\omega t&\omega x^2\cos\omega t&\omega x^3\cos\omega t\\
\chi\sin\omega t&\cos\omega t&\omega x^1\sin\omega t&\omega x^2\sin\omega t&\omega x^3\sin\omega t\\
\omega x^1&0&1+\lambda {x^1}^2&\lambda x^1 x^2&\lambda x^1 x^3\\
\omega x^2&0&\lambda x^1 x^2&1+\lambda {x^2}^2&\lambda x^2 x^3\\
\omega x^3&0&\lambda x^1 x^3&\lambda x^2 x^3&1+\lambda {x^3}^2
\end{array}\right)\,,\nonumber\\
\end{eqnarray} 
where we denote $\lambda=\frac{\chi -1}{r^2}$. Indeed, calculating $z(x)= {\frak b}(t,\vec{x}) z_o$ we obtain just  Eqs. (\ref{Zx}). Moreover, we may derive the canonical tetrads  defined by Eq. (\ref{Om}) obtaining the non-vanishing components
\begin{eqnarray}
&\hat e^{0}_0=\chi\,,&\quad  \hat e^{i}_j=\delta^i_j-\frac{x^i x^j}{r^2}\left(1-\frac{1}{\chi}\right)\,,\\
& e^{0}_0=\frac{1}{\chi}\,,&\quad   e^{i}_j=\delta^i_j-\frac{x^i x^j}{r^2}\left(1-{\chi}\right)\,.
\end{eqnarray}
We recover thus the Cartesian gauge \cite{ES} which preserves the global $SO(3)$ symmetry.  

Other charts can be introduced in the same manner by using suitable boosts. For defining the coordinates of the chart  $\{\tilde x\}$ we chose the boost  
\begin{eqnarray}
&&{\frak b}(t,\rho,\theta,\phi)=\exp\left(-i {\frak J}_3\, \phi \right)\exp\left(-i {\frak J}_2\, (\theta-\frac{\pi}{2}) \right)\nonumber\\
&& \times \exp\left(-i {\frak H}\,\omega t\right) \exp\left(-i {\frak N}_1 \,\beta\right)\exp\left(\frac{i\pi}{2} {\frak J}_1\right)\,,\\
&&\beta={\rm arctanh}\, \omega\rho\,,
\end{eqnarray}
where  the last rotation is a gauge transformation determining convenient 1-forms. After a little calculation we obtain
\begin{eqnarray}
&&{\frak b}(t,\rho,\theta,\phi)=\nonumber\\
&&\left(
\begin{array}{ccccc}
\frac{1}{\tilde\chi}\cos\omega t&-\sin\omega t&\frac{\omega\rho}{\tilde\chi}\cos\omega t&0&0\\
\frac{1}{\tilde\chi}\sin\omega t&\cos\omega t&\frac{\omega\rho}{\tilde\chi}\sin\omega t&0&0\\
\frac{\omega\rho}{\tilde\chi}\sin\theta\cos\phi&0&\frac{1}{\tilde\chi}\sin\theta\cos\phi&\cos\theta\cos\phi&-\sin\phi\\
\frac{\omega\rho}{\tilde\chi}\sin\theta\sin\phi&0&\frac{1}{\tilde\chi}\sin\theta\sin\phi&\cos\theta\sin\phi&\cos\phi\\
\frac{\omega\rho}{\tilde\chi}\cos\theta&0&\frac{1}{\tilde\chi}\cos\theta&-\sin\theta&0
\end{array}\right)\,,\nonumber\\
\end{eqnarray}
and we may convince ourselves that $z(\tilde x)={\frak b}(t,\rho,\theta,\phi)z_o$ gives the 
embedding equations (\ref{Ztx}). In this case the canonical 1-forms read
\begin{eqnarray}
\hat\omega^0&=&\frac{1}{\tilde\chi(\rho)}dt\,,\\
\hat\omega^1&=&\frac{1}{\tilde\chi(\rho)^2}d\rho\,,\\
\hat\omega^2&=&\frac{\rho}{\tilde\chi(\rho)}d\theta\,,\\
\hat\omega^3&=&\frac{\rho\sin\theta}{\tilde\chi(\rho)}d\phi\,,
\end{eqnarray}
defining a familiar diagonal gauge. 

\subsection{Lorentzian isometries}

Let us consider now the problem of the relative motion that studies how a geodesic motion can be measured by two different observers. The above introduced local charts play the role of inertial frames related through isometries. Each observer may have a proper frame in which he stays at rest in origin on the world line along the vector field $\partial_t$. In general, these frames move along geodesics such that we need to introduce supplemental hypotheses in order to describe their motion. 

We assume that one observer, $O$, is fixed in his proper frame $\{x\}$ observing what happens in a mobile frame $\{x'\}$ which is the proper frame of the observer $O'$. The problem is how this relative motion can be described using isometries. We start with the hypothesis that the mobile frame is the rest frame of a particle of mass $m$ 
which stays at rest in $O'$ having $E'=m$ and ${\vec{L}}'={\vec{K}}'={\vec{N}}'=0$. The fixed observer $O$ see this particle moving on a timelike geodesic with given parameters that mark the relative motion. Under such circumstances, the previously presented boosting method allows us to derive the isometry transformation between these frames in terms of the geodesic parameters related with the conserved quantities. 

This relativity does make sense only if we can compare the measurements of these observers imposing the condition of the synchronization of their clocks. This means that, at a given common initial time, the origins of these frames must coincide.  However, this condition is restrictive in the CAdS spacetimes since this forces the geodesic of the carrier particle of the mobile frame  to across the origin of the fixed frame $O$. This  means that its trajectory is rectilinear (with $\vec{L}=0$)  in a given direction $\vec{n}$ on which this oscillates with frequency $\omega$.  

With these preparations we can apply the boosting method in order to relate the five-dimensional vectors of $M^5$ corresponding to the fixed and mobile frames. In the mobile frame the observer $O'$ stays at rest in ${\vec{x}\,}'=0$ measuring the time $t'$. There are no restrictions to choose an arbitrary time $t'_0$  of $O'$ giving the point    
\begin{equation}
z'(x_0')=\left(\frac{1}{\omega}\cos\omega t'_0,\frac{1}{\omega}\sin\omega t'_0, 0,0,0\right)^T \in M^5\,,
\end{equation} 
while in the fixed frame the corresponding point 
 \begin{equation}
z(x_0)=\left(\frac{1}{\omega}\cos\omega t_0,\frac{1}{\omega}\sin\omega t_0, x^1_0,x^2_0,x^3_0\right)^T \in M^5\,,
\end{equation} 
depends on the position $\vec{x}_0$ at time $t_0$ of the particle of mass $m$. The boosts (\ref{Cboost}) allows us to reach both these points starting from $z_o$ as, 
$z(x)={\frak b}(t,\vec{x})z_o$ and $z'(x')={\frak b}(t',0)z_o$. This gives us the opportunity of defining  ${\frak g}_*= {\frak b}(t_0,\vec{x}_0){\frak b}(t_0',0)^{-1}\in SO(2,3)$ which relates the above defined points in $M^5$ by the transformation  $z(x_0)={\frak g}_*z'(x'_0)$. 

Furthermore, we assume that the isometry transformation between the frames $\{x\}$ and $\{x'\}$ is generated by ${\frak g}_*$. However, the synchronization condition requires the origins to coincide for $t=t'=0$ when we must have $\vec{x}={\vec{x}\,}=0$  such that $O$ and $O'$ overlap in the same point $z(0)=z'(0)=z_o\in M^5$. This happens only if we chose an initial condition of type (II.b), given by Eq. (\ref{IIb}), setting 
\begin{equation}\label{tt00}
t_0=t_0'=\frac{1}{2}\frac{\pi}{\omega}\,.
\end{equation}
Thus we obtain the desired transformation
\begin{equation}\label{gst}
{\frak g}_*=\left(
\begin{array}{ccccc}
1&0&0&0&0\\
0&\chi_0&\omega n^1r_0&\omega n^2 r_0&\omega n^3 r_0\\
0&\omega n^1 r_0&1+\lambda'{n^1}^2&\lambda' n^1n^2&\lambda' n^1n^3\\
0&\omega n^2 r_0&\lambda'n^1n^2&1+\lambda'{n^2}^2&\lambda'n^2n^3\\
0&\omega n^3 r_0&\lambda'n^1n^3&\lambda'n^2n^3&1+\lambda'{n^3}^2
\end{array}\right)\,,
\end{equation}
resulted after  we substituted $\vec{x}_0=\vec{n} r_0$, laying out the unit vector $\vec{n}$ fixing the direction of the  linear trajectory and the {amplitude} $r_0$ in the chart $\{t,\vec{x}\}$ that gives $\chi_0=\sqrt{1+\omega^2 r_0^2}$ and  $\lambda'=\chi_0-1$.

It is remarkable that ${\frak g}_*$  is in fact a genuine Lorentz transformation of the $SO(2,3)$ group, having the form ${\frak g}_*=\exp\left(-i {\frak K}_i n^i \sigma \right)$ where
\begin{equation}
\sigma= {\rm arctanh}\left(\frac{\omega r_0}{\chi_0}\right)={\rm arctanh}(\omega\rho_0)\,.
\end{equation}
Thus we obtained a Lorentz transformation even though we started with boosts generated by ${\frak N}_i$. The explanation is that by imposing the condition (\ref{tt00}) we perform the transformation 
\begin{equation}
\exp\left(-\,\frac{i \pi}{2}{\frak H}\right) {\frak N}_i \exp\left(\,\frac{i \pi}{2}{\frak H}\right)={\frak K}_i
\end{equation} 
changing the generators of ${\frak g}_*$.

The transformation ${\frak g}_*$ allows us to find the conserved quantities on the geodesic of the particle of mass $m$. Taking into account that in its rest frame $O'$ the mobile has $E'=m$ and ${\vec{L}}'={\vec{K}}'={\vec{N}}'=0$, corresponding to the matrix ${\cal K}_o$, we may apply the general rule for finding the conserved quantities observed by $O$, encapsulated in the matrix  ${\cal K}=\overline{\frak g}_*\,{\cal K}_o\,\overline{\frak g}_*^T$.  After a little calculation we obtain  $\vec{L}=\vec{K}=0$, as expected, and the non-vanishing conserved quantities $E=m\chi_0$ and $\vec{N}=m \vec{x}_0$ with the help of which  we  can express all the other constants as 
\begin{equation}\label{chro}
\chi_0=\frac{E}{m} \to r_0=\frac{\sqrt{E^2-m^2}}{\omega m} \to \rho_0=a=\frac{\sqrt{E^2-m^2}}{\omega E}\,.
\end{equation} 
These formulas suggest us to define a momentum vector
\begin{equation}\label{mom}
\vec{P}=\omega \vec{N}=m\omega \vec{x}_0=\vec{n} P\,, \quad P=\sqrt{E^2-m^2}\,,
\end{equation}
which satisfy formally the usual dispersion relation of special relativity and whose meaning will be discussed later. Now we use this vector for writing down the definitive expression of the Lorentz transformation (\ref{gst}) as
\begin{equation}\label{gfin}
{\frak g}_*={\frak g}(\vec{P})=\exp\left( -i\vec{P}\cdot\vec{{\frak K}}\,\frac{1}{P}\, {\rm arcsinh}\left(\frac{P}{m}\right)   \right)\,,
\end{equation}
which can be obtained by substituting $\chi_0=\frac{E}{m}$ and $r_0=\frac{P}{m}$ in Eq. (\ref{gst}). We observe that this matrix has the same form as the corresponding Lorentz boost of special relativity \cite{WKT} or as the transformation giving the Lorenzian isometry of the dS relativity  \cite{CdS}.

The last step is to define the Lorenzian isometry $x=\phi_{{\frak g}(\vec{P})}(x')$, between the coordinates of the mobile and fixed frames, as the transformation resulted from the system of equations $z(x)={{\frak g}_*}z'(x')$ that can be derived using the new parametrization  (\ref{gfin}). The direct transformation reads
\begin{eqnarray}
t(t',{\vec{x}\,}')&=&\frac{1}{\omega}\,{\rm arctan}\left(\frac{E}{m}\tan\omega t'\right. \nonumber\\
&&\left.+\frac{\omega}{m}\,\frac{{\vec{x}\,}'\cdot \vec{P}}{\sqrt{1+\omega^2 |\vec{x}\,'|^2}} \sec\omega t'\  \right)\,,\label{Lort}\\
\vec{x}(t',{\vec{x}\,}')&=&{\vec{x}\,}'+\frac{\vec{P}}{m}\left(\frac{{\vec{x}\,}'\cdot\vec{P}}{E+m}\right.\nonumber\\
&&\left.+\frac{1}{\omega}\sqrt{1+\omega^2 |\vec{x}\,'|^2}\sin\omega t'\right)\,,\label{Lox}
\end{eqnarray}
while the inverse one has to be obtained by changing $x \leftrightarrow x'$ and $\vec{P}\to -\vec{P}$. We obtained thus the CAdS analogous of the Lorentz isometries of special relativity.   Obviously, in the limit of $\omega \to 0$ we recover the usual Lorentz transformations of special relativity. 

The first application of this isometry  is to recover the geodesic trajectory of the carrier particle of mass $m$  from the parametric equations in $t'$ obtained by substituting ${\vec{x}\,}'=0$ in Eqs. (\ref{Lort}) and (\ref{Lox}). Then, according to Eqs. (\ref{chro}) and (\ref{mom}), we obtain the trajectory of the origin $O'$ denoted as
\begin{eqnarray}
\vec{x}_*(t)=\vec{n}r_*(t)&=&\vec{n}\frac{a\sin\omega(t)}{\sqrt{1-\omega^2 a^2 \sin^2 \omega (t)}}\nonumber\\
&=&\frac{\vec{P}\sin(\omega t)}{\omega \sqrt{E^2-P^2\sin^2\omega t}}\,,
\end{eqnarray} 
which corresponds to our initial condition $\vec{x}_*(0)=0$.   The components of the four-velocity,   
\begin{eqnarray}
u^0_*(t)&=&\frac{d t}{ds}=\frac{m^2+P^2\cos^2\omega t}{m E}\,,\\
 \vec{u}_*(t)&=&\frac{d \vec{x}_*(t)}{ds}=\frac{E \vec{P}\cos(\omega t)}{m\sqrt{m^2+P^2\cos^2\omega t}}\,,
\end{eqnarray}
show that the momentum we introduced above is just the canonical momentum of the mobile $m$ at $t=0$ since $E=mu_*^0(0)$ and  $\vec{P}=m \vec{u}_*(0)$.  

Now we may use as principal parameter the velocity of the carrier particle at $t=0$, defined usually as $\vec{V}=\frac{\vec{P}}{E}=\vec{n} V$,  if we desire to bring the above formulas in  forms closer to those of special relativity, eliminating the mass $m$.  This can be done by changing the parametrization,  
\begin{equation}
\frac{E}{m}=\gamma\,,\quad \frac{P}{m}=\gamma V\,,\quad  \gamma=\frac{1}{\sqrt{1-V^2}}\,,
\end{equation}  
and substituting   $\chi_0=\gamma$ and $r_0=\frac{\gamma V}{\omega}$ in Eq. (\ref{gst}). Then we can rewrite
\begin{equation}\label{gfin1}
{\frak g}(\vec{P}) \to{\frak g}(\vec{V})=\exp\left( -i\vec{V}\cdot\vec{\frak K}\,\frac{1}{V}\, {\rm arctanh}\left(V\right)   \right)\,,
\end{equation}
obtaining the new expression of the Lorentzian isometry 
\begin{eqnarray}
t(t',{\vec{x}\,}')&=&\frac{1}{\omega}\,{\rm arctan}\,\gamma \left(\tan\omega t'\right.\nonumber\\
&&\left.+\,\frac{\omega{\vec{x}\,}'\cdot \vec{V}}{\sqrt{1+\omega^2 |\vec{x}\,'|^2}} \sec\omega t'\  \right)\,,\label{Lort}\\
\vec{x}(t',{\vec{x}\,}')&=&{\vec{x}\,}'+\gamma \vec{V}\left({\vec{x}\,}'\cdot\vec{V}\frac{\gamma}{1+\gamma}\right.\nonumber\\
&&\left.+\frac{1}{\omega}\sqrt{1+\omega^2 |\vec{x}\,'|^2}\sin\omega t'\right)\,,\label{Lox}
\end{eqnarray}
that may be used in applications. 

The transformations ${\frak g}(\vec{V})$ generating these isometries, transform simultaneously all the conserved quantities. If those of the mobile frame are encapsulated in the matrix ${\cal K}'$ as in Eq. (\ref{KK}), then the corresponding ones measured in the fixed frame are the matrix elements of the matrix
\begin{equation}\label{KKK}
{\cal K}=\overline{\frak g}(\vec{V}) \,{\cal K}'\, \overline{\frak g}(\vec{V})^T\,.
\end{equation}
Thus we obtain the principal tools in studying the relative motion on CAdS spacetimes, or simply, CAdS relativity.

\section{Relativistic effects}

This approach gives us the opportunity of solving various applications. In what follows we present some interesting consequences that can be deduced from Eqs. (\ref{Lort}), (\ref{Lox}) and (\ref{KKK}) but avoiding the complicated technical details.

\subsection{Time dilation and Lorentz contraction}  

The simplest  interesting effects are the time dilation (observed in the so called twin paradox) and the Lorentz contraction which in this case are quite complicated since the these effects are strongly dependent on the position where the time and length are measured. For this reason,  we restrict ourselves to give a mere simple example, assuming  that the measurements are performed in a small neighborhood  of ${\vec{x}}'=0$.   Here we consider the general relations 
\begin{eqnarray}
\delta t &=&\left.\frac{\partial t(t',{\vec{x}}')}{\partial t'}\right| _{{\vec{x}}'=0}\delta t' + \left.\frac{\partial t(t',{\vec{x}}')}{\partial x^{\prime\, i}}\right| _{{\vec{x}}'=0}\delta  x^{\prime\, i}\,,\\
\delta x^j &=&\left.\frac{\partial x^j(t',{\vec{x}}')}{\partial t'}\right| _{{\vec{x}}'=0}\delta t' +\left.\frac{\partial x^j(t',{\vec{x}}')}{\partial x^{\prime\, i}}\right| _{{\vec{x}}'=0}\delta  x^{\prime\, i}\,,
\end{eqnarray} 
among the quantities $\delta t, \delta x^j$ and $\delta t',\delta x^{\prime\, j}$ supposed to be measured by the observers $O$ and respectively $O'$.  

{ \begin{figure}
    \centering
    \includegraphics[scale=.65]{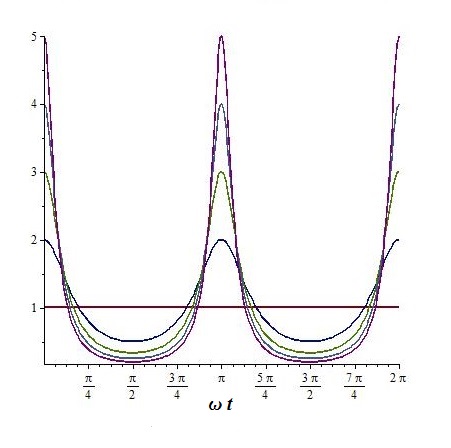}
    \caption{The function $\tilde\gamma(t)$ for different values, $\gamma=1,\,\gamma=2,\,...\gamma=5$.  }    
  \end{figure}}
  
First we consider a clock in $O'$ indicating $\delta t'$ without changing its position such that $\delta x^{\prime\, i}=0$. Then, after a little calculation, we obtain the time dilation observed by $O$,    
\begin{equation}
\delta t=\delta t' \tilde\gamma(t)\,,\quad \tilde\gamma(t)=\frac{\gamma}{\gamma^2 \sin^2\omega t +\cos^2\omega t}\,.
\end{equation}   
Similarly but with the supplemental simultaneity condition $\delta t=0$ we derive the Lorentz contraction along the direction of $\vec{P}$ that reads
\begin{equation}
\delta x_{||}=\delta x'_{||}\frac{1}{\tilde\gamma(t)}\,.
\end{equation}   
It is remarkable that here we have $\delta t \delta x_{||}=\delta t' \delta x'_{||}$ just as in the flat case. 

The difference is that now the function $\tilde\gamma(t)$ is oscillating in the domain $[\frac{1}{\gamma}, \gamma]$, producing a time dilation for $\tilde\gamma>1$ and a time contraction if $\tilde\gamma<1$, during the same period (as in Fig. 4). This example shows how interesting may be the kinematics of the free motion on the CAdS spacetime. 

\subsection{Relative geodesic motion}

However, the principal challenge of  relativity is to find how an arbitrary geodesic motion with respect to the mobile frame can be measured by the fixed observer $O$. Let us consider the mobile frame $\{x'\}$ as the local chart of Cartesian space coordinates defined with respect to the orthogonal basis $\{{\vec{\nu}\,}_1', {\vec{\nu}\,}'_2,{\vec{\nu}\,}'_3  \}$,  assuming that its origin has the velocity $\vec{V}$ when is passing through the origin of a fixed frame.   In this mobile frame, a test particle of mass $m$ moves in the plane $\{ {\vec{\nu}\,}'_1, {\vec{\nu}\,}'_2\}$ on an ellipsoidal geodesic trajectory whose major axis is oriented along the unit vector ${\vec{\nu}\,}'_1$,  obeying equations of the form (\ref{geo1})-(\ref{geo3})  that now read
\begin{eqnarray}
{x'}\,^1(t')&=&\frac{\rho'_{\rm max} }{\sqrt{1-\omega^2 \rho'(t')^2}}\,\cos \omega (t'-t_0')\,,\label{geo1a} \\
{x'}\,^2(t')&=&\frac{\rho'_{\rm min} }{\sqrt{1-\omega^2 \rho'(t')^2}}\,\sin \omega (t'-t'_0)\,,\label{geo2a}\\
{x'}\,^3(t')&=&0\,,\label{geo3a}
\end{eqnarray}
where, for simplicity, we choose the initial condition $t_0'=0$ (when the test particle reaches the aphelion in the mobile frame). The trajectory parameters in the mobile frames are determined by the energy $E'$ and  angular momentum $\vec{L}\,'=(0,0,L')$ of the test  particle. Thus, we may calculate $\rho'_{\rm max}$ and $\rho'_{\rm min}$  according to  Eqs. (\ref{rhomm}),  (\ref{Kap1}) and (\ref{Kap2}), while the other conserved quantities result from Eqs. (\ref{KER1})-(\ref{NER2}) as
\begin{equation}\label{roro}
\vec{K}'=(E'\rho_{\rm max}',0,0)\,, \quad \vec{N}'=(0,-E'\rho_{\rm min}',0)\,,
\end{equation}
since $t'_0=0$. Under such circumstances, the fixed observer $O$ measures the geodesic trajectory of the test particle resulted after applying the Lorentzian transformation, substituting Eqs. (\ref{geo1a})-{\ref{geo3a}) in Eqs. (\ref{Lort})-(\ref{Lox}).  In this manner, we obtain  the parametric equations of the geodesic trajectory in the fixed frame,
\begin{eqnarray}
t(t')&=& \frac{1}{\omega}\arctan \gamma \left[ \omega\rho'_{\rm max}V^1\right.\nonumber\\
&&\left.~~~~~~~~~~~~~~~+\left(1+\omega\rho_{\rm min}'V^2 \right)\tan\omega t'\right]\,,\\
x^1(t')&=&\frac{1}{\sqrt{1-\omega^2\rho'(t')^2}}\left[\left( 1+(V^1)^2\frac{\gamma^2}{1+\gamma}\right)\rho'_{\rm max}\cos\omega t'\right.\nonumber\\
&&\left.~~~~~+\frac{\gamma}{\omega}V^1\left(1+\omega\rho_{\rm min}'V^2\frac{\gamma}{1+\gamma}\right)\sin\omega t'\right]\,,\\
x^2(t')&=&\frac{1}{\sqrt{1-\omega^2\rho'(t')^2}}\left[V^1V^2\frac{\gamma^2}{1+\gamma}\rho'_{\rm max}\cos\omega t'+\left(\frac{\gamma}{\omega}V^2\right.\right.\nonumber\\
&&\left.\left. +\left(1+(V^2)^2\frac{\gamma^2}{1+\gamma}\right)\rho_{\rm min}'\right)\sin\omega t'\right]\,,\\
x^3(t')&=&\frac{1}{\sqrt{1-\omega^2\rho'(t')^2}}\left[V^1V^3\frac{\gamma^2}{1+\gamma}\rho'_{\rm max}\cos\omega t'\right.\nonumber\\
&&~~~~~~~~\left.\frac{\gamma}{\omega}V^3\left( 1+V^2\frac{\gamma}{1+\gamma}\omega\rho_{\rm min}'\right)\sin\omega t'\right]\,.
\end{eqnarray}
depending on the parameter $t'$ that can be eliminated for deriving the equations $\vec{x}(t)$ of the geodesic of the test particle in this frame. 

The basis $\{{\vec{\nu}\,}_1', {\vec{\nu}\,}'_2,{\vec{\nu}\,}'_3  \}$ which is fixed rigidly  in the mobile frame, determining its Cartesian space coordinates,  is seen by the fixed observer as a moving basis $\{ {\vec{\nu}_1}(t), {\vec{\nu}_2}(t),{\vec{\nu}_3}(t)  \}$.  The time dependence of this basis can be found by using the Lorentzian isometry. We assume that each unit vector ${\vec{\nu}\,}'_i$ is observed at its own time $t'_i$, such that we may impose the simultaneity condition 
\begin{equation}
t=t[t_1', \vec{\nu}_1\,']=t[t_2', \vec{\nu}_2\,']=t[t_3', \vec{\nu}_3\,']\,,
\end{equation}
giving the functions $t'_i(t)$ we need for writing down the form of the unit vectors of the mobile basis as
\begin{equation}
{\vec{\nu}\,}_i (t)=\vec{x}[t'_i(t),{\vec{\nu}\,}'_i], \quad i=1,2,3\,.
\end{equation}

The last task is to find the conserved quantities measured by the fixed observer as resulted from the transformation (\ref{KKK}). Let us consider the general case of an arbitrary geodesic orbit of a test particle in the mobile frame whose kinematics is determined by the conserved quantities. The problem is how many such  independent quantities we need for determining the kinematic parameters. We observe that six parameters, say the components of $\vec{K}$ and $\vec{N}$, are enough for finding the energy and angular momentum components from Eqs. 
(\ref{invariant}) and (\ref{property}b). Moreover, Eqs. (\ref{prop1})-(\ref{prop3}) give the trajectory parameters, including the time $t_0$ when the test particle is passing through the aphelion. 

However, here we consider seven parameters, the above mentioned vectors and, in addition, the energy $E'$ instead of the mass $m$, since thus we avoid the difficulties arising when one uses explicitly Eq. (\ref{invariant}). Therefore,  starting with $E'$, $\vec{K}'$ and $\vec{N}'$ in the mobile frame and applying the transformation (\ref{KKK}), we derive the general rules:  
\begin{eqnarray}
E&=&\gamma \left(E'+\omega\, \vec{V}\cdot\vec{N}'\right)\,,\\
\vec{K}&=&\gamma\vec{K}'+\frac{\omega\gamma}{E'}\left[\vec{K}'(\vec{V}\cdot\vec{N}')-\vec{N}'(\vec{V}\cdot\vec{K}')) \right]\nonumber\\
&& -\frac{\gamma^2}{1+\gamma}\vec{V}(\vec{V}\cdot\vec{K}')\,,\\
\vec{N}&=&\vec{N}'+\frac{\gamma}{\omega}E'\vec{V}+\frac{\gamma^2}{1+\gamma}\vec{V}(\vec{V}\cdot\vec{N}')\,,
\end{eqnarray}
giving the conserved quantities measured in the fixed frame where the angular momentum  results from Eq. (\ref{property}b) as
\begin{equation}
\vec{L}=-\frac{\omega}{E}\, \vec{K}\land\vec{N}\,.
\end{equation}
One can verify that this transformation preserves the invariant (\ref{invariant}) which defines the mass of the test particle. Thus the problem of  CAdS relativity is completely solved in terms of conserved quantities.

\section{Concluding remarks}

We presented here the principal methods of studying the relative geodesic motion on CAdS 
spacetimes, based on the $SO(2,3)$ transformation generating the Lorenzian isometry that may be expressed in terms of conserved  parameters. Starting with this transformation,  we derived the general rule of transforming the conserved quantities from the mobile natural frame to the fixed one, obtaining thus information about all the kinematic parameters we need without  resorting to the complicated algebraic calculations of the geodesic equations in a particular local chart.  

Technically speaking the CAdS relativity is somewhat similar with the dS one \cite{CdS} since both these spacetimes are hyperboloidal and can be constructed as spaces of left cossets on the gauge groups of their embedding flat manifolds. However, the similarity stops here since their embedding manifolds and isometry groups are different, determining different types of geodesic trajectories or quantum modes.  Thus on  dS spacetimes the geodesics are open and the quantum modes correspond to continuous energy spectra as it happens in the Minkowski spacetime too. On the contrary, on the AdS spacetimes the geodesic motion is oscillatory and the quantum modes are square-integrable wave functions corresponding to equidistant discrete energy spectra with the same quanta $\hbar \omega$ (in IS units) regardless the spin  \cite{Cq1,Cq2}.

Finally, we note that the embedding method may be used for solving  common dS-CAdS problems, in an $(2+4)$-dimensional  common embedding flat spacetime whose gauge group is just the conformal one, $SO(2,4)$. In this geometry we may study  new problems as, for example, what happens in a CAdS sphere moving on the dS spacetime.  We hope that our methods presented in Ref. \cite{CdS} and here will be useful for solving such new problems in an  extended (special) relativity of hyperbolic spacetimes, considering all the fifteen conserved quantities corresponding to the $SO(2,4)$ symmetry. 

\appendix

\section{Inverse problem}

There are situations when we know the integration constants $\kappa_1$, $\kappa_2$, $\phi_0$ and $t_0$ and we need to find the physical conserved quantities. Then from Eqs. (\ref{Kap1}) and (\ref{Kap3}) we deduce
\begin{eqnarray}
E&=&\frac{m}{\sqrt{(\kappa_1\omega^2-1)^2-\kappa_2^2\omega^4}}\,,\\
L&=&\frac{m\omega\sqrt{\kappa_1^2-\kappa_2^2}}{\sqrt{(\kappa_1\omega^2-1)^2-\kappa_2^2\omega^4}}\,.
\end{eqnarray}  
Furthermore, from Eqs. (\ref{K1})-(\ref{N2}), calculated in aphelion, at $t=t_0$, where $\tilde u^{\rho}=0$ and $\tilde u^t$ and $\tilde u^{\phi}$ result from Eqs. (\ref{conE}) and (\ref{conL}), we derive the non-vanishing components
\begin{eqnarray}
K_1&=&E\rho_{\rm max}\cos\phi_0\cos\omega t_0+E\rho_{\rm min}\sin\phi_0\sin\omega t_0\,,\\
K_2&=&E\rho_{\rm max}\sin\phi_0\cos\omega t_0-E\rho_{\rm min}\cos\phi_0\sin\omega t_0\\
N_1&=&-E\rho_{\rm max}\cos\phi_0\sin\omega t_0+E\rho_{\rm min}\sin\phi_0\cos\omega t_0\,,\\
N_2&=&-E\rho_{\rm max}\sin\phi_0\sin\omega t_0-E\rho_{\rm min}\cos\phi_0\cos\omega t_0\,.
\end{eqnarray}
while $K_3=N_3=0$. These components satisfy the properties  (\ref{prop1})-(\ref{prop3}).


\begin{thebibliography}{20}

\bibitem{W}
S. Weinberg, {\it Gravitation and Cosmology: Principles and Applications of 
the General Theory of Relativity}, (Wiley, New York, 1972).

\bibitem{Cal}
E. Calabi and L. Markus,  {\em Ann. of Mathematics} {\bf 75}, 63 (1962). 

\bibitem{Haw}
S. W. Hawking and G.F. R. Ellis,  {\em  The large scale structure of space-time},
(Cambridge Univ. Press, Cambridge, 1973).

\bibitem{Grif}
J. Griffiths and J. Podolsky, {\em Exact space-times in Einstein's general
relativity}, (Cambridge Univ. Press, Cambridge, 2009). 

\bibitem{P1}
E. van Beveren, G. Rupp, T. A. Rijken and C. Dullemond, {\it Phys. Rev.} 
{\bf D27}, 1527 (1983)

\bibitem{P2}
C. Dullemond and E. van Beveren, {\it Phys. Rev.} {\bf D28}, 1028 (1983)

\bibitem{L}
A. R. Lugo, hep-th/9904163.

\bibitem{C1}
I. I. Cotaescu and D. N. Vulcanov, {\em EPL} {\bf 49}, 156  (2000).

\bibitem{SG}
L. M. Sokołowski  and Z. A. Golda, {\em Int.  J.  Mod. Phys. D} {\bf  25}, 1650007 (2016).

\bibitem{Biz}
L. M. Sokolowski, {\em Int. J. of Geometric Methods in Mod. Phys.} {\bf 13}, 1630016 (2016).

\bibitem{Mal}
J. Maldacena, {\em Adv. Theor. Math. Phys.} {\bf 2}, 231 (1998).

\bibitem{AdSCFT}
A review and extended bibliography in: V. E. Hubeny, arXiv:1501.00007v2.

\bibitem{ES}
I. I. Cot\u aescu, {\em J. Phys. A: Math. Gen.} {\bf 33}, 9177  (2000).

\bibitem{CGRG}
I. I. Cot\u aescu, {\em GRG} {\bf 43}, 1639 (2011).

\bibitem{CdS}
I. I. Cot\u aescu, {\em Eur. Phys. J. C} {\bf 77}, 485 (2017); arXiv:1701.08499. 

\bibitem{CAdS}
I. I. Cot\u aescu, {\em Phys. Rev. D} {\bf 95}, 104051 (2017); arXiv:1702.06835.

\bibitem{Nach}
O. Nachtmann, {\em Commun. Math. Phys.} {\bf 6}, 1 (1967).

\bibitem{Wig}
E. Wigner, {\em Ann. Math.} {\bf 40}, 149  (1939). 

\bibitem{WKT}
W.-K. Tung,  {\em Group Theory in Physics}  (World Sci., Philadelphia, 1984).

\bibitem{Cq1}
I. I. Cot\u aescu, {\em Phys. Rev. D} {\bf  60}, 107504 (1999).

\bibitem{Cq2}
I. I. Cot\u aescu, {\em Phys. Rev. D} {\bf 60}, 124006 (1999).

\end{thebibliography}
\end{document}